\documentclass[12pt,preprint]{emulateapj}

\usepackage{graphicx}
\usepackage{epsfig}
\usepackage{subfigure}

\def \aap{A\&A }
\def \apjl{ApJ}

\def \apj{ApJ}

\def \mnras{MNRAS}
\def \prl{Phys.~Rev.~Lett.,}

\def \nat{Nature\ }
\def\msun{{\,M_\odot}}

\def\alt{\raise0.3ex\hbox{$\;<$\kern-0.75em\raise-1.1ex\hbox{$\sim\;$}}}
\def\agt{\raise0.3ex\hbox{$\;>$\kern-0.75em\raise-1.1ex\hbox{$\sim\;$}}}

\newcommand{\HI}{H{\sc i}}

\newcommand{\bw}{\begin{widetext}}
\newcommand{\ew}{\end{widetext}}

\newcommand{\lsim}{\,\rlap{\raise 0.35ex\hbox{$<$}}{\lower 0.7ex\hbox{$\sim$}}\,}
\newcommand{\gsim}{\,\rlap{\raise 0.35ex\hbox{$>$}}{\lower 0.7ex\hbox{$\sim$}}\,}

\begin{document}

%

\title{Steady-State Hadronic Gamma-ray Emission from 100-Myr-Old Fermi Bubbles}

\author{\sc
  Roland M.~Crocker\altaffilmark{1,3},
  Geoffrey V.~Bicknell\altaffilmark{1}, Ettore Carretti\altaffilmark{2}, Alex S.~Hill\altaffilmark{2}, Ralph S.~Sutherland\altaffilmark{1}
}
\altaffiltext{1}{Research School of Astronomy and Astrophysics, Australian National University, Canberra, Australia}
\altaffiltext{2}{CSIRO Astronomy \& Space Science, Marsfield, N.S.W., Australia}
\altaffiltext{3}{Future Fellow}

\date{\today}

\begin{abstract}

The Fermi Bubbles are enigmatic $\gamma$-ray features of the Galactic bulge.
Both putative activity (within few $\times$ Myr) connected to the Galactic center super-massive black hole
and, alternatively, nuclear star formation have been claimed as the energising source of the Bubbles.
Likewise, both  inverse-Compton emission by non-thermal electrons (`leptonic' models) and  collisions between non-thermal protons and gas (`hadronic' models)
have been advanced as the  process supplying the Bubbles' $\gamma$-ray emission.
An issue for any steady state hadronic model is that the very low density of the Bubbles' plasma seems to require that they accumulate protons over a multi-Gyr timescale, 
much longer than other natural timescales occurring in the problem. 
Here we present a  mechanism 
wherein the  timescale for generating the Bubbles'  $\gamma$-ray emission via hadronic processes is $\sim$ few $\times 10^8$ years.
Our model invokes collapse of the Bubbles' thermally-unstable plasma, leading to an accumulation of cosmic rays and magnetic field 
into localised, warm ($\sim 10^4$ K), and likely filamentary condensations of higher density gas.
Under the condition that these filaments are supported by non-thermal pressure,  the hadronic emission from the Bubbles is
$L_\gamma \simeq 2 \times 10^\mathrm{37}$ erg/s$ \ \dot{M}_\mathrm{in}/(0.1 \ \msun/$year$) \ T_\mathrm{FB}^2/(3.5 \times 10^7$ K)$^2 \ M_{fil}/M_{pls}$,
equal to their observed luminosity (normalizing to the star-formation-driven mass flux into the Bubbles and their measured plasma temperature and adopting the further result 
that the mass in the filaments, $M_{fil}$ is approximately equal to the that of the Bubbles' plasma,  $M_{pls}$).
%
%

\end{abstract}

\maketitle

\section{Introduction}

The Fermi Bubbles  \citep{Dobler2009,Su2010} are giant 1-100 GeV $\gamma$-ray structures that extend $\sim$7 kpc north and south  from the Galactic nucleus.
Structures, roughly coincident with the Bubbles, are known in 
X-rays \citep{Bland-Hawthorn2003},
total intensity microwave \citep{Finkbeiner2004,Dobler2008,Dobler2011,Ade2012}, polarised intensity microwave \citep{Jones2012}, and polarised intensity radio continuum \citep{Carretti2013} emission.

Much theoretical work on the Bubbles has focused on the idea that their $\gamma$-ray emission is supplied by the inverse-Compton (IC) up-scattering 
of ambient light by a cosmic ray (CR) electron population.
Given that the Bubbles extend so far into the halo 
with a relatively fixed, hard spectrum, an IC model must invoke up-scattering of the CMB 
to multi-GeV gamma-ray energies, requiring $\gsim$TeV electron primaries.
Such electrons cool quickly defining a natural timescale   $\lesssim 1$ Myr.
To explain, then, the large extent and energy content of the Bubbles some models \citep[e.g.,][]{Zubovas2012,Guo2012,Yang2012,Yang2013,Barkov2013} 
hypothesise recent activity of the super-massive black hole at the Galactic Center (GC) 
with very fast transport of the electrons.
These timing constraints are relaxed if there is distributed acceleration \citep{Cheng2011,Mertsch2011,Lacki2013} throughout the structures.
Evidence in support of an AGN-like scenario may come from the detection \citep{Su2012} of a jet-like feature in the $\gamma$-ray data 
and the recent claim that the Magellanic stream was bathed in a bright UV flash only a few million years ago \citep{Bland-Hawthorn2013}.

Alternatively,   the Bubbles' $\gamma$-ray emission may be hadronic in origin with  CR protons 
(and heavier
ions) 
-- ultimately energised by nuclear star-formation -- accumulated over (much) longer timescales,
colliding with ambient gas to supply the $\gamma$-rays  \citep{Crocker2011}.
A number of pieces of evidence are consistent with this scenario.
First,  
the intense star-formation in the inner $\sim$200 pc diameter region around the Galactic center
{\it currently} produces a cosmic ray power   
that is elegantly sufficient to supply both the (hadronic) $\gamma$-ray luminosity of the Bubbles  \citep{Crocker2011b,Crocker2012}
and the 2.3 GHz synchrotron luminosity of the  recently-discovered, 
polarised radio lobe counterparts to the Bubbles \citep{Carretti2013}.
Second, 
as we  show below \citep[also see][]{Crocker2012}
the mass flux of $\sim 0.1 \msun$/year from the nuclear-star-formation driven outflow is
elegantly sufficient to maintain the
Bubbles' plasma mass in steady-state against its thermal losses.
Last and more speculatively, in the sky distribution of
 up to 28 neutrino events recently observed by IceCube against a  background of 10.6 atmospheric events, there is a hint of an overabundance of neutrinos
 from the direction of the Inner Galaxy with an inferred flux consistent with a hadronic origin for the  $\gamma$-rays \citep{Aartsen2013,Ahlers2013,Razzaque2013}.

The apparent cost, however, of  any steady state hadronic model 
is the long timescale implied.
The protons and ions fed into the Bubbles collide with ambient gas nuclei over the $pp$ loss time which,
on the low volumetric average gas density, is $t_\mathrm{pp} \simeq$ 10 Gyr $[n_H/(0.005 $ cm$^\mathrm{-3})]^\mathrm{-1}$.
To establish a steady state requires that the structures have existed for this sort of timeframe, difficult to reconcile with other natural timescales relating to the Bubbles/GC.
In particular, the  2.3 GHz polarisation observations \citep{Carretti2013} suggest the outflow feeding 
electrons into the radio lobes has a vertical speed of $\sim 1100$  km/s giving an advective timescale over the $\sim$8 kpc extent of the radio lobes 
(marginally larger than the Bubbles) 
of only $7 \times 10^6$ yr.  
Nuclear star-formation, which supplies a
total mechanical power of $\sim 3 \times 10^{40}$ erg/s, $\sim 80$\% of which goes into heating or moving the outflowing plasma \citep{Crocker2012},
 cannot supply all the Bubbles' energy ($\gsim 5 \times 10^\mathrm{55}$ erg) within this timescale.

We  show immediately below, however, that both this short advection timescale and the long timescale associated with steady-state hadronic emission need to be reconsidered
given a natural mechanism that is the subject of this {\it Letter}.
Briefly, this is the following: cooling of the Bubbles' interior plasma over $1-2 \times 10^8$ years into cool, over-dense filaments leads to the adiabatic compression of cosmic rays and magnetic fields.
With the simple prescription that the (dominantly non-thermal) pressure of the filaments reaches equilibrium with the Bubble plasma, we can predict their hadronic emission.
As we show below, this prediction is a good match to observations

\section{Timescales Reconsidered}

The $\sim 1100$ km/s characteristic speed at the base of the outflow is suggested by two independent analyses of the radio data based on: 
i) a geometrical analysis of the curvature of one of the strongly magnetised, linear `ridge' features seen in the data and 
ii)  the cooling time of the electrons whose synchrotron emission reaches to the top of the radio lobes.
This speed is somewhat in excess of the gravitational escape velocity \citep[$\sim 900$ km/s;][]{Muno2004}  and, if interpreted as purely a plasma velocity, 
 expected to imply an escaping wind. 
%
%

In fact, however,  the fast flow {\it does not escape to infinity}:
the 23 GHz polarisation data show that the magnetic field lines  curl over to  near horizontal at the top of the Bubbles,  suggesting a vertical speed that is significantly slowing.
Thus there is a structure of a definite, finite height,  not a  freely expanding wind.
The CRs stream ahead of the plasma with a speed limited to the Alfven velocity because of the streaming instability.
This velocity also controls the speed with which torsional Alfven waves respond to the driving from the rotation of the outflow's base.
We thus  re-interpret the previously-determined 1100 km/s 
characteristic speed as the sum of two components: a plasma wind speed {\it that is somewhat less than the escape velocity} 
(but still capable of delivering gas to $\sim$ few kpc heights into the halo)
and the Alfven speed along the magnetic field lines.
Thus
 material and energy can be accumulated into the Bubbles for significantly longer than $7 \times 10^6$ yr.

The second problem  is  to reconcile the very long hadronic timescale with other timescales.
For instance, the thermally unstable Bubble plasma  cools over $(1-2) \times 10^8$ yr (see below), collapsing into over-dense condensations.
This concentrates the magnetic field attached to the plasma, thereby mustering in the CRs.
Analogous to the processes occurring around the Brightest Central Galaxies (BCGs) of clusters \citep{Sharma2010}, this collapse will continue until the non-thermal pressure in the condensations equilibrates with the  external plasma pressure.
The CRs and magnetic fields evolve adiabatically at the beginning of the collapse, in contrast to the plasma whose collisional losses rise strongly as $n^2$.
Thus the non-thermal ISM phases come to dominate the pressure in the condensations despite their softer equation of state.
As for the H$\alpha$ filaments around BCGs \citep{Sharma2010}, 
anisotropic heat conduction (due to suppression of electron motions transverse to magnetic fields) means that 
the condensations  assume the topology of extended filaments with dominantly longitudinal  magnetic fields.
With this picture, we can calculate how the collapse proceeds.

\section{Hadronic Emission from Cooling Filaments}

The Milky Way has an X-ray bulge (XRB) with some individual features  clearly correlated with Bubble structures \citep{Su2010}.
The {\it bolometric} thermal luminosity of the XRB lies in the range $L_\mathrm{XRB} = (3-9) \times 10^\mathrm{39}$ erg/s \citep{Snowden1997,Almy2000}.
From this luminosity and assuming a static structure we can calculate the rate at which material is cooling out of the  XRB as $\dot{M}_\mathrm{cool} \simeq 2/3 \ L_\mathrm{XRB} \ \mu m_\mathrm{p}/(k_\textrm{B} T) \sim (0.07 - 0.2) \msun$/year ($\mu \simeq 0.6 $ is the mean mass of the plasma constituents in terms of $m_\mathrm{p}$).
This $\dot{M}_\mathrm{cool}$ is  close to the  mass efflux \citep{Crocker2012} along the SF-driven nuclear outflow.
Given that the mechanical energy {\it and} mass injected by nuclear SF can sustain the non-thermal luminosity of the Bubbles in steady state, we postulate that the Bubbles are enduring structures inflated and sustained by nuclear star formation.
In this circumstance  freshly-injected plasma from the nuclear outflow balances the mass drop-out rate, $\dot{M}_\mathrm{in} \equiv \dot{M}_\mathrm{cool}$, resulting in a steady state gas density in the Bubbles of
\begin{eqnarray}
n_\mathrm{H^+}^\mathrm{FB} &=& \left(\frac{\dot{M}_\mathrm{in}  \ T_\mathrm{FB} \ k_\textrm{\tiny{B}}  }{(\gamma-1) \ \mu \ 1.22 \ m_\mathrm{p} \ V_\mathrm{FB} \ \Lambda[T]} \right)^\mathrm{1/2} \nonumber \\ 
&\simeq &0.003 \ \textrm{cm}^\mathrm{-3} \left(\frac{\dot{M}_\mathrm{in}}{0.1 \msun/\textrm{year}} \right)^\mathrm{1/2}
\label{eqn_1}
\end{eqnarray}
where  
 $n_\mathrm{e} \simeq 1.22 n_\mathrm{H^+}$, $(\gamma-1) = 2/3$, $\Lambda[T]$ is the plasma thermal cooling function \citep{Raymond1976}, and $V_\mathrm{FB} = 8.4 \times 10^{66}$ cm$^3$.
Reinforcing the steady state picture, this $n_\mathrm{H^+}^\mathrm{FB}$ estimate is consistent with X-ray measurements by  SUZAKU  \citep{Kataoka2013} 
which give  
$T_\mathrm{FB} \simeq 3.5 \times 10^6$ K for the Bubble plasma 
(equal to that of the adjacent halo plasma)
 and
from which we infer $n_\mathrm{H^+}^\mathrm{FB} \simeq (1-3) \times 10^\mathrm{-3}$ cm$^\mathrm{-3}$. 

For this plasma number density and temperature the cooling time is $(1-2) \times 10^8$ year.
We show below that the timescale for the formation of the Bubbles is $\gsim 2 \times 10^8$ years,
only a little longer.

Given adiabatic compression of relativistic CR protons into the plasma condensations, 
we can make an estimate of the hadronic $\gamma$-ray emission from the Bubbles, finding consistency with   observations.
The filament pressure, supplied by relativistic CR protons (denoted by $p$) 
and magnetic fields ($B$), equilibrates with the external plasma pressure, $p^\mathrm{fil} \equiv p_\mathrm{p}^\mathrm{fil} + p_B^\mathrm{fil} = p^\mathrm{FB} = 2.2 \times T_\mathrm{FB} \ k_\mathrm{\tiny{B}} \ n_\mathrm{H^+}^\mathrm{FB} = 2.0$ eV cm$^\mathrm{-3}$ and we suppose   $p_\mathrm{p}^\mathrm{fil}  \simeq p_B^\mathrm{fil} $.
%
%
The energy density of the adiabatically-accumulated, relativistic CR protons in the filaments is $u_p^\mathrm{fil} = u_p^\mathrm{GC} (n_\mathrm{H}^\mathrm{fil}/n_\mathrm{H}^\mathrm{GC})^\mathrm{4/3} $ where GC 
denotes a parameter of the nuclear star formation region where the CRs and thermal plasma are energised.
With $p_p^\mathrm{fil} \equiv 1/3 \ u_p^\mathrm{fil} = 1/2 \ p^\mathrm{fil}$ we determine the  filament filling factor as $f = (3.3 \ k_\textrm{B} \ T_\mathrm{FB} n_\mathrm{H^+}^\mathrm{FB}/u_p^\mathrm{GC})^\mathrm{3/4} \ n_\mathrm{H^+}^\mathrm{FB}/n_\mathrm{H^+}^\mathrm{GC}$
and the hadronic luminosity from 1-100 GeV of the filaments is
\begin{eqnarray}
L_\gamma^\mathrm{pp} &\simeq& 3/2 \ 1/3 \ f_\mathrm{bolo} \ \sigma_\mathrm{pp} \ \kappa_\mathrm{pp} \ c \ n_\mathrm{H}^\mathrm{fil} \ u_\mathrm{p}^\mathrm{fil} \  \ V_\mathrm{fil} \nonumber \\
&\simeq& \frac{3  \ f_\mathrm{bolo} \ \sigma_\mathrm{pp} \ \kappa_\mathrm{pp} \ c \ \dot{M}_\mathrm{in} (k_\mathrm{\tiny{B}}  \ T_\mathrm{FB})^2}{\Lambda[T_\mathrm{FB}] \ m_\mathrm{p}}   
 \left( \frac{M_{fil}}{M_{pls}} \right)  \\
&\simeq& 2 \times 10^\mathrm{37} \textrm{erg/s} \left(\frac{\dot{M}_\mathrm{in}}{\scriptscriptstyle{0.1  \msun/\textrm{\tiny{yr}}}} \right) \left(\frac{T_\mathrm{FB}}{\scriptscriptstyle{3.5 \times 10^6 {\textrm \tiny{K}}}} \right)^2  \left( \frac{M_{fil}}{M_{pls}} \right)  \nonumber
\end{eqnarray}
where $\kappa_\mathrm{pp} \simeq 0.5$ is the inelasticity of $pp$ collisions, a factor of 3/2  corrects for the presence of heavy ions  \citep{Mori1997}, $1/3$ comes from the relative multiplicity of $\pi^0$ amongst all daughter pions, and $f_\mathrm{bolo} \simeq 0.4$ is the fraction of the bolometric luminosity emitted in the 1-100 GeV range.
(As in the original hadronic model of \citet{Crocker2011}, this scenario reproduces the Bubbles' hard spectrum given the energy-independence of the relevant transport processes 
 and near energy-independence of $\sigma_\mathrm{pp}$ above the kinematic threshold.)
Thus, normalizing to the measured plasma temperature and the mass injection rate required to maintain its  density in steady state (Eq.~\ref{eqn_1}),
the expected luminosity matches that  observed \citep{Lunardini2012}  {\it provided} that the filaments' integrated mass, $M_{fil}$,  approximately equals that in the plasma 
$M_{pls} (\ \simeq 2 \times 10^7 \msun)$. 

Remarkably, this condition is met:
given the steady state, we have $M_{fil}/M_{pls} \equiv \langle t_\textrm{\tiny{fall}} \rangle /t_\textrm{\tiny{cool}}$ with $\langle t_\textrm{\tiny{fall}} \rangle$
the mean time for the filaments
to fall to the plane {\it at their terminal speed} \citep[cf.][]{Benjamin1997} through the Bubble plasma of density $\rho_\textrm{\tiny{pls}}$.
For a horizontally-aligned filament this is given by
\begin{equation}
v_\mathrm{term}^\textrm{horiz}[r,z] \simeq \left( \frac{\pi}{2}  \frac{\rho_\textrm{\tiny{fil}}}{\rho_\textrm{\tiny{pls}}}    \frac{r_\textrm{\tiny{fil}} \ g[r,z]}{c_D}  \right)^{1/2}
\end{equation}
where $g[r,z]$ is the magnitude of the gravitational acceleration at $\{r,z\}$ and $c_D$ is the drag coefficient 
(for a vertically-aligned filament of length $L_\textrm{\tiny{fil}}$ replace $ \pi/2\  r_\textrm{\tiny{fil}} \to L_\textrm{\tiny{fil}})$.
The filament terminal velocity is super-Alfvenic through the Bubble plasma; this implies draping of the Bubbles'
magnetic field around the filaments, the formation of magnetic wakes behind the filaments,
and a consequent increase in drag with respect to the  hydrodynamic expectation that is accounted for by setting $c_D \simeq 1.9$  which adopt here \citep[][]{Dursi2008}.
Employing the potential described by \citet{Breitschwerdt1991}, 
we find the mean time for the filaments to fall from their condensation sites at $z_\textrm{\tiny{launch}}$
 to the plane at the terminal speed to be
$t_\textrm{\tiny{fall}}[z] \simeq 8 \times 10^7 \ (z_\textrm{\tiny{launch}}/$4 kpc)$^{0.7}$ year.
Accounting for the fact that filaments do not form below $\sim 3$ kpc (see below), we find that the mean filament falling time satisfies
 $\langle t_\textrm{\tiny{fall}} \rangle \sim t_\textrm{\tiny{cool}}$
to better than a factor 2.
This agreement is not accidental but essentially guaranteed by the following considerations:
While buoyancy effects mean that a hot, gravitationally-confined and stratified atmosphere is {\it not} generally susceptible to the local cooling instability
\citep{Balbus1989,Binney2009},
it has been empirically and theoretically established  \citep{McCourt2012,Sharma2012,Li2013} that cooling filaments {\it can} form 
within a medium in global thermal balance
wherever the ratio of the
cooling to the {\it free} fall time, $t_\mathrm{ff}$,  satisfies $t_\mathrm{cool}/t_\mathrm{ff} < (3-10)$, particularly if 
the medium is subject to external perturbations.
This timescale ratio condition is satisfied in the Bubbles for $z \gsim 3$ kpc.
Thus, because  $t_\textrm{\tiny{fall}}  \sim$ few $\times t_\mathrm{ff}$, we automatically have $t_\textrm{\tiny{cool}}/t_\textrm{\tiny{fall}} \lsim$ few in any environment where filaments form.

\section{Discussion}
      
At equilibrium the filaments are compressed to  $f \sim10$\% and
  $n_H^\mathrm{fil} \sim 0.03$ cm$^\mathrm{-3}$.
At any given filament $n_\mathrm{H^+}$ we require that 
the filaments are warm enough to obey the (conservative) H$\alpha$ intensity upper limit $\sim$ 0.3 Rayleigh \citep{Finkbeiner2004};
this condition means that the filament gas  must be warmer than $\sim 5000$ K.
At this temperature their cooling rate is $3 \times 10^\mathrm{39}$ erg/s, $\sim$10\% of the mechanical energy injected into the nuclear outflow.
The filament temperature may thus be maintained by an internal agent associated with the outflow
-- shock heating, thermal conduction, dissipation of hydrodynamical turbulence, magnetic field reconnection, and/or CR excitation of MHD waves or direct CR ionisation --
or an external one, e.g., photoionisation heating from Lyman continuum photons supplied by the young nuclear stars.

The filament filling factor we favour, $\sim$
10\%, corresponds to a layer of thickness $\sim 100$ pc if distributed evenly over the (assumed) spherical volume of each bubble.
The compressed magnetic field amplitude is expected to be $\sim 10 \ \mu$G.
These numbers are a near match to the depth (200-300 pc) and amplitude ($\sim 10 \ \mu$G) for the magnetised sheath 
suggested to cover the Bubbles by \citet{Carretti2013}.

Within our model, CRs are initially distributed throughout the volume of the Bubbles and are subsequently gathered in to the filaments  by the collapse of the thermally-unstable plasma.
This process requires that the  inward convective velocity  exceeds the effective outward  velocity associated with CRs' lateral escape.
Because the filaments' field lines are largely longitudinal, lateral escape is
 via cross-field diffusion or, more importantly, field line wandering \citep{Jokipii1969}.
In either case, lateral escape is generically 
much slower than diffusion {\it along} the filaments which is diffusive with diffusion coefficient $D_\mathrm{||}$, but limited to the Alfven velocity because of the streaming instability.
In the case of field line wandering, the expectation value of the square of the perpendicular distance reached in time $t$ is given by $\langle r^2 \rangle = 4 \ D_M \sqrt{2 \ D_\mathrm{||} \ t}$ \citep{Nava2013}. 
$D_M$, the diffusion coefficient for the field lines (with dimensions of length), is poorly constrained.
 If we adopt $D_M = 1$ pc from \citet{Nava2013} and a parallel diffusion coefficient $D_\mathrm{||}$ similar to that of the Galactic  plane, 
 we determine that CR protons up to an energy $\sim 1$ TeV 
 are trapped within the filaments over the $\sim 10^8$ year cooling time if their  inward transversal collapse (down to final radius $r_f$) proceeds at 
 $\sim \sqrt{f} \ r_f/t_\mathrm{cool} \sim 3$ km/s $ r_f/(100$ pc) $(t_\mathrm{cool}/(10^8$ year)$^{-1}$.
 With these parameter choices, higher energy CR protons start to escape the filaments within the cooling timescale; consistent with this, 
 there is a steepening in the Bubbles' $\gamma$-ray spectrum at $\sim 100$ GeV, corresponding to primary protons of energy $\sim 1 $ TeV.

We suggest an intimate connection between the Bubbles and the nuclear molecular torus, 
which is fed by the Galactic bar and akin to nuclear star-forming rings found in other barred spirals.
The torus constitutes much of the mass of the CMZ \citep{Molinari2011} and, with a gravitational potential energy of few $\times 10^\mathrm{56}$ erg, 
it is the logical candidate to anchor the  Bubbles' field lines.
The natural timescale associated with the formation of the  torus  is also $\gsim 10^8$ years over which it has hosted the formation of $\sim 10^7 \msun$ of stars and consequently $\sim 10^5$ core-collapse supernovae that have released $\sim 10^\mathrm{56}$ erg mechanical energy.
The torus seems \citep{Zubovas2012,Crocker2012}, moreover, to collimate the GC outflow and material ablated off its inner edge can naturally
supply the  \HI \ clouds recently detected \citep{McClure-Griffiths2013} at relatively low latitudes, $|b| < 5^\circ$,  entrained into a nuclear outflow.
These $n_\mathrm{HI}
 \sim 1$ cm$^\mathrm{-3}$ clouds 
 will help nucleate the condensation of plasma \citep{Marinacci2010}, 
 especially towards the edges of the Bubbles.
To 
match their flat $\gamma$-ray surface brightness   a volumetric emissivity 
that strongly peaks towards the edges of the Bubbles is required \citep{Su2010}; 
filament nucleation occurring preferentially near the edges may achieve this.

Direct evidence for warm, ionised gas with characteristics very similar to those we infer for the filaments has been obtained via UV absorption studies \citep{Keeney2006,Zech2008}.
These works have uncovered warm, ionised gas clouds high above and below the GC with number densities $\lsim$ few $\times \ 0.1$ cm$^\mathrm{-3}$ and temperatures in the range few $\times \ 10^4$ K.
These clouds have insufficient velocity to escape the Galaxy and are thus participating in a nuclear fountain.
Tantalizingly, all the  sight lines that reveal fountaining warm plasma  are within the solid angle of the Bubbles or the somewhat larger radio lobes.
Of particular interest,  the sightline to the Messier 5 globular cluster \citep{Zech2008} passes very close to the edge of the north Bubble and reveals
 super-solar metallicity ($\sim 1.6 Z_\odot $) fountain material, consistent with the scenario of filament
 nucleation occurring preferentially towards the Bubble edges.

While we believe that nuclear star formation ultimately powers the  Bubbles, 
the mechanism of CR hadron accumulation into cooling filaments 
does not necessarily require that the CRs are accelerated (only) in the nucleus.
They may, for instance, be (re)accelerated on significantly larger scales by shocks in the outflow \citep[cf.][]{Lacki2013}.
Indeed, our analysis would still hold in the case that
an outburst from Sgr A$^*$ $\gsim 10^8$ years ago inflated the Bubbles. 
Regardless, the fountaining-back of relatively cool and low angular momentum filament gas to the plane may occasionally provide a cold accretion flow  on to the black hole or fuel star-formation very close to it.


RMC is the recipient of an Australian Research Council Future Fellowship (FT110100108). 
RMC gratefully acknowledges conversation or correspondence with Felix Aharonian, Joss Bland-Hawthorn, Vladimir Dogiel,  Stefano Gabici, Ortwin Gerhard, Randy Jokipii, Brian Lacki, Federico Marinacci, 
Mark Morris, Martin Pohl, Vladimir Ptuskin, Mateuz Ruszkowski, Prateek Sharma, Meng Su, and Daniel Wang.
%



\end{document}